\documentclass[aps,onecolumn,qic,superscriptaddress,11pt]{revtex4-1}

\usepackage[T1]{fontenc}
\usepackage[sc]{mathpazo}
\usepackage{amsmath}
\usepackage{amssymb}
\usepackage{enumerate}
\usepackage{amsthm}% theorems and proofs
\usepackage{color}

\usepackage{amsfonts,mathrsfs}
\usepackage[style]{fncychap}
\usepackage{graphicx} % For including graphics N.B. pdftex graphics driver
\usepackage{geometry} % allows easy specifying of the page layout
\usepackage{eepic}
\usepackage{ifthen}
\newboolean{ElectronicVersion}
\setboolean{ElectronicVersion}{true} % CHANGE THIS AS REQUIRED

\usepackage[T1]{fontenc}
\usepackage[sc]{mathpazo}
\usepackage{mathdots}
\usepackage{amsmath}
\usepackage{amssymb}
\usepackage{enumerate}
\usepackage{amsthm}% theorems and proofs
\usepackage{amsfonts,mathrsfs}
\usepackage[style]{fncychap}
\usepackage{graphicx} % For including graphics N.B. pdftex graphics driver
\usepackage{geometry} % allows easy specifying of the page layout
\usepackage{eepic}
\usepackage{ifthen}
\usepackage{cases}

\usepackage{amsfonts,mathrsfs}
\usepackage[style]{fncychap}
\usepackage{graphicx} % For including graphics N.B. pdftex graphics driver
\usepackage{geometry} % allows easy specifying of the page layout
\usepackage{eepic}
\usepackage{ifthen}
\usepackage{cases}

\usepackage{hyperref}
\hypersetup{pdfpagemode=UseNone}

\begin{document}\pagestyle{empty}

\title{Reply to \\ Comment on ``Multiparty quantum mutual information: An
alternative definition''}

\author{Asutosh Kumar}
\email{asutoshk.phys@gmail.com}

%\affiliation{Vaidic and Modern Physics Research Centre, Bhagal Bhim, Bhinmal, Jalore 343029, India}
 \affiliation{Department of Physics, Gaya College, Magadh University, Rampur, Gaya 823001, India}

\begin{abstract}
We reaffirm the claim of Lee {\it et al.} [preceding Comment, Phys. Rev. A {\bf 108}, 066401 (2023)] that the expression of quantum dual total correlation of a multipartite system in terms of quantum relative entropy as proposed in previous work [A. Kumar, Phys. Rev. A {\bf 96}, 012332 (2017)] is not correct. We provide alternate expression(s) of quantum dual total correlation in terms of quantum relative entropy. We, however, prescribe that in computing quantum dual total correlation one should use its expression in terms of von Neumann entropy.  
\end{abstract}

\maketitle

\section{Introduction}
In Ref. \cite{asu-qmi} two different expressions of quantum dual total correlation were obtained: one in terms of von Neumann entropy and other in terms of quantum relative entropy. It was claimed that the two expressions are equivalent.
In a comment \cite{lee} on Ref. \cite{asu-qmi}, Lee {\it et al.} have shown that the quantum dual total correlation of an $n$-partite quantum state cannot be represented as the quantum relative entropy between $(n - 1)$ copies of the quantum state and the product of $n$ different reduced quantum states for $n \ge 3$.
They arrived at this conclusion by considering explicitly the ``support'' condition of quantum relative entropy. 
Essentially, what Lee {\it et al.} have shown is that the following two expressions are not equal for $n \ge 3$:

\begin{equation} \label{vn-ent}
I_n(\rho) := \sum^n_{k=1} S(\rho_{\overline{k}}) - (n-1) S(\rho),
\end{equation}
where $\rho_{\overline{k}} = tr_k (\rho)$ denotes the $(n-1)$-partite quantum state obtained by taking the partial trace on the $k^{th}$ party of $\rho$, and 

\begin{equation}\label{qr-ent1}
J_n(\rho) := S(\rho^{\otimes (n-1)} || \otimes^n_{k=1} \rho_{\overline{k}}),
\end{equation}
where the quantum relative entropy is 
%$S(\rho || \sigma) := tr \rho (\log \rho - \log \sigma)$ if $supp(\rho) \subseteq supp(\sigma)$, and infinity otherwise. 
\[
 S(\tau || \sigma) := 
  \begin{cases} 
   tr (\tau \log \tau) - tr (\tau \log \sigma) & \text{if } supp(\tau) \subseteq supp(\sigma) \\
   \infty       & \text{otherwise.} 
  \end{cases}
\]

To justify their claim, authors provide two examples which imply that the above two expressions of $n$-partite quantum mutual information are not equivalent. 
%These two examples are simple and clear. 
$I_n(\rho)$  in Eq. (\ref{vn-ent}) is non-negative and non-increasing under local CPTP maps \cite{cerf}, and therefore is a suitable monotonic measure of multi-partite correlations, while $J_n(\rho)$ in Eq. (\ref{qr-ent1}) is not. 

\section{Reaffirming claim of Lee {\it et al.}}
The claim of Lee {\it et al.} is right. 
In this article we show analytically why the above two expressions are not equivalent. We begin with expression of $I_n(\rho)$ [Eq. (\ref{vn-ent})] and proceed to show that this is not equal to $J_n(\rho)$ [Eq. (\ref{qr-ent1})], as argued below.

\begin{eqnarray}
I_n(\rho) &=& \sum^n_{k=1} S(\rho_{\overline{k}}) - (n-1) S(\rho) \nonumber \\
%&=& \sum^n_{k=1} S(\rho_{\overline{k}}) + \sum^n_{k=1} S(\rho_{k}) - n S(\rho) +S(\rho) - \sum^n_{k=1} S(\rho_{k}) \nonumber \\
&=& \sum^n_{k=1} \big(S(\rho_{k}) + S(\rho_{\overline{k}}) - S(\rho) \big) - \big(\sum^n_{k=1} S(\rho_{k}) - S(\rho) \big) \nonumber \\
&=& \sum^n_{k=1} S(\rho || \rho_k \otimes \rho_{\overline{k}}) - S(\rho || \otimes^n_{k=1} \rho_{k}) \label{qmi1} \\
&=& S \big(\rho^{\otimes n} || \otimes^n_{k=1} (\rho_{k} \otimes \rho_{\overline{k}}) \big) - S(\rho || \otimes^n_{k=1} \rho_{k}) \label{qmi2} \\
&\overset{?}{=}& {\color{blue}S \big(\rho \otimes \rho^{\otimes (n-1)} || (\otimes^n_{k=1}~ \rho_{k}) \otimes (\otimes^n_{k=1}~ \rho_{\overline{k}}) \big)} - S(\rho || \otimes^n_{k=1} \rho_{k}) \label{problem} \\
&=& S(\rho || \otimes^n_{k=1} \rho_{k}) + S(\rho^{\otimes (n-1)} || \otimes^n_{k=1} \rho_{\overline{k}}) - S(\rho || \otimes^n_{k=1} \rho_{k}) \nonumber \\
&=& S(\rho^{\otimes (n-1)} || \otimes^n_{k=1} \rho_{\overline{k}}) = J_n(\rho), \label{qmi3}
\end{eqnarray}
where 
%$S(\rho || \rho_k \otimes \rho_{\overline{k}})$ 
quantum relative entropy in Eq. (\ref{qmi1}) and Eq. (\ref{qmi2}) is properly matched to satisfy the ``support'' condition in the sense that 
\begin{eqnarray}
S(\rho || \otimes^n_{k=1} \rho_{k}) &\equiv & S(\rho_{12\cdots n} || \otimes^n_{k=1} \rho_{k}) \nonumber \\
S(\rho || \rho_k \otimes \rho_{\overline{k}}) &\equiv & S(\rho_{k\overline{k}} || \rho_k \otimes \rho_{\overline{k}}) \nonumber \\
S \big(\rho^{\otimes n} || \otimes^n_{k=1} (\rho_{k} \otimes \rho_{\overline{k}}) \big) &\equiv & S \big(\rho_{12\cdots n} \otimes \rho_{23\cdots n1} \otimes \cdots \otimes \rho_{n1\cdots (n-1)} || \otimes^n_{k=1} (\rho_{k} \otimes \rho_{\overline{k}}) \big), \nonumber
\end{eqnarray}
where $\overline{1} = 23 \cdots n$, $\overline{2} = 34 \cdots n1$ and $\overline{k} = (k+1) \cdots n1 \cdots (k-1)$. Eq. (\ref{qmi1}) and Eq. (\ref{qmi2}) are alternate expressions equivalent to Eq. (\ref{vn-ent}) in terms of quantum relative entropy. We, however, prescribe to use Eq. (\ref{vn-ent}) in computing quantum dual total correlation.
%Because tensor product is not commutative, Eq. (\ref{problem}) is not correct and hence we cannot arrive at Eq. (\ref{qmi3}). 
Eq. (\ref{problem}) is not correct for two reasons: (i) noncommutativity of tensor product, and (ii) ``matching'' issue of subsystems. Therefore,  we cannot arrive at Eq. (\ref{qmi3}). 

\section{Second Rebuttal}
Let us reconsider Eq. (\ref{qr-ent1}) for $n=3$ explicitly.
\begin{eqnarray}
J_3(\rho) &=& S(\rho^{\otimes 2} || \otimes^3_{k=1} \rho_{\overline{k}}) \nonumber \\ 
&=& S(\rho_{123} \otimes \rho_{123} || \rho_{23} \otimes \rho_{31} \otimes \rho_{12}). \label{match1} 
\end{eqnarray}
Here we see that the subsystems in the first and the second arguments of the quantum relative entropy are not properly matched. 
However, the subsystems in the above expression could be matched up if we adopt the following conventions: (i) interpret the first argument in the usual way, with the subsystems in their standard order and (ii) interpret the tensor product in the second argument with the values of $k$ running from $n$ to $1$. That is, for $n=3$, if we define 
\begin{eqnarray}
\tilde{J}_3(\rho) &:=& S(\rho^{\otimes 2} || \otimes^1_{k=3} \rho_{\overline{k}}) \nonumber \\
&=& S(\rho_{123} \otimes \rho_{123} || \rho_{12} \otimes \rho_{31} \otimes \rho_{23}), \label{match2} 
\end{eqnarray}
then we see that the subsystems are properly matched. Now, let us adopt the following notations:
\begin{eqnarray}
\rho_{123}^{\otimes 2} &=& \rho_{123} \otimes \rho_{123} 
 \equiv  \rho_{A_1A_2A_3} \otimes \rho_{B_1B_2B_3} 
 \equiv  \rho_{123} \otimes \rho_{456}, \label{notation1} \\
\otimes^1_{k=3} \rho_{\overline{k}} &=& \rho_{12} \otimes \rho_{31} \otimes \rho_{23} 
 \equiv  \rho_{A_1A_2} \otimes \rho_{A_3B_1} \otimes \rho_{B_2B_3} 
 \equiv  \rho_{12} \otimes \rho_{34} \otimes \rho_{56}. \label{notation2}
\end{eqnarray}
Then, using notations in Eqs. (\ref{notation1}, \ref{notation2}), one might attempt to show that Eq. (\ref{match2}) is equivalent to $I_3(\rho) = \sum^3_{k=1} S(\rho_{\overline{k}}) - 2 S(\rho)$ as argued below.
\begin{eqnarray}
\tilde{J}_3(\rho) &:=& S(\rho^{\otimes 2} || \otimes^1_{k=3} \rho_{\overline{k}}) \nonumber \\
&=& S(\rho_{123} \otimes \rho_{123} || \rho_{12} \otimes \rho_{31} \otimes \rho_{23}) \nonumber \\
&\equiv & S(\rho_{123} \otimes \rho_{456} || \rho_{12} \otimes \rho_{34} \otimes \rho_{56}) \nonumber \\
&=& - tr (\rho_{123} \otimes \rho_{456} \log (\rho_{12} \otimes \rho_{34} \otimes \rho_{56})) + tr (\rho_{123} \otimes \rho_{456} \log (\rho_{123} \otimes \rho_{456})) \nonumber \\
&=& - tr (\rho_{123} \otimes \rho_{456} \log (\rho_{12} \otimes I_{34} \otimes I_{56})) - tr (\rho_{123} \otimes \rho_{456} \log (I_{12} \otimes \rho_{34} \otimes I_{56})) \nonumber \\
&-& tr (\rho_{123} \otimes \rho_{456} \log (I_{12} \otimes I_{34} \otimes \rho_{56})) + tr (\rho_{123} \otimes \rho_{456} \log (\rho_{123} \otimes I_{456})) + tr (\rho_{123} \otimes \rho_{456} \log (I_{123} \otimes \rho_{456}))\nonumber \\
&=& - tr_{12} (\rho_{12} \log \rho_{12}) {\color{blue} - tr_{34} (\rho_{3} \otimes \rho_{4} \log \rho_{34})} - tr_{56} (\rho_{56} \log \rho_{56}) + tr_{123} (\rho_{123} \log \rho_{123}) + tr_{456} (\rho_{456} \log \rho_{456}) \nonumber \\
&\overset{?}{=}& S(\rho_{12}) + {\color{blue}S(\rho_{34})} + S(\rho_{56}) - S(\rho_{123}) - S(\rho_{456}) \label{qmi4} \\
&\equiv & S(\rho_{12}) + S(\rho_{31}) + S(\rho_{23}) - 2 S(\rho_{123}) = I_3 (\rho).
\end{eqnarray}
But this is not correct because the second term in Eq. (\ref{qmi4}) cannot be obtained from the previous equation.

Thus, even if we define quantum dual total correlation of a multipartite system in terms of quantum relative entropy alternatively as
\begin{equation}\label{qr-ent2}
\tilde{J}_n(\rho) := S(\rho_{12\cdots n}^{\otimes (n-1)} || \otimes^1_{k=n} \rho_{\overline{k}}),
\end{equation}
and use the notations as discussed above, Eq. (\ref{qr-ent2}) is not equivalent to Eq. (\ref{vn-ent}).

\begin{acknowledgments}
AK is very thankful to Jaehak Lee, Gibeom Noh, Changsuk Noh and Jiyong Park for pointing out the subtle mistake in Ref. \cite{asu-qmi}. 
%Any comment on this article is welcome. 
\end{acknowledgments}

\end{document}